\documentstyle[aps,prl,preprint,floats,epsfig]{revtex}

\def\rwsmeas   {\relax\ifmmode{0.0043^{+0.0011}_{-0.0010}\ ({\rm stat}) \pm
    0.0007\ ({\rm syst})} 
  \else{$0.0043^{+0.0011}_{-0.0010}\ ({\rm stat}) \pm
    0.0007\ ({\rm syst})$}\fi}
\def\asymmnumber {\relax\ifmmode{A=(8^{+25}_{-22})\%} 
         \else{$A=(9^{+25}_{-22})\%$}\fi}

\def\rwsmeasstat   {\relax\ifmmode{0.0043^{+0.0011}_{-0.0010}\ ({\rm stat})} 
  \else{$0.0043^{+0.0011}_{-0.0010}\ ({\rm stat})$}\fi}
\def\signifsig {\relax\ifmmode{4.9}
                 \else{$4.9$}\fi}
\def\nwsevnt {\relax\ifmmode{38\pm 9}
                 \else{$38\pm 9$}\fi}
\def\nrsevnt {\relax\ifmmode{8837\pm xx}
                 \else{$8837\pm xx$}\fi}
\def\rhopctyld{\relax\ifmmode{79\%}
                 \else{$79\%$}\fi}
\def\ksppctyld{\relax\ifmmode{16\%}
                 \else{$16\%$}\fi}
\def\kszpctyld{\relax\ifmmode{13\%}
                 \else{$13\%$}\fi}
\def\allrhoerr{\relax\ifmmode{2\%}
                 \else{$2\%$}\fi}
\def\allksperr{\relax\ifmmode{1\%}
                 \else{$1\%$}\fi}
\def\allkszerr{\relax\ifmmode{28\%}
                 \else{$28\%$}\fi}

\def\totalsyserr{\relax\ifmmode{17\%}
                 \else{$17\%$}\fi}
\def\errdalbgdst{\relax\ifmmode{3\%}
                 \else{$3\%$}\fi}
\def\errdalfitstat{\relax\ifmmode{2\%}
                 \else{$2\%$}\fi}
\def\errQmshapes{\relax\ifmmode{14\%}
                 \else{$14\%$}\fi}
\def\onehyperr{\relax\ifmmode{3.6\%}
                 \else{$3.6\%$}\fi}
\def\effratiototerr{\relax\ifmmode{9.5\%}
                 \else{$9.5\%$}\fi}
\def\mccutdisterr{\relax\ifmmode{3\%}
                 \else{$3\%$}\fi}
\def\errotherres{\relax\ifmmode{8\%}
                 \else{$8\%$}\fi}
\def\errmcstat{\relax\ifmmode{2.4\%}
                 \else{$2.4\%$}\fi}
\def\errchrmfrag{\relax\ifmmode{0\%}
                 \else{$0\%$}\fi}
\def\dzdzb{\relax\ifmmode{D^{0}-\overline{D^{0}}}
                 \else{$D^{0}-\overline{D^{0}}$}\fi}
\def\kzkzb{\relax\ifmmode{K^{0}-\overline{K^{0}}}
                 \else{$K^{0}-\overline{K^{0}}$}\fi}
\def\bzbzb{\relax\ifmmode{B^{0}-\overline{B^{0}}}
                 \else{$B^{0}-\overline{B^{0}}$}\fi}
\def\dztokppzws{\relax\ifmmode{D^{0}\rightarrow K^{+}\pi^{-}\pi^{0}}
                 \else{$D^{0}\rightarrow K^{+}\pi^{-}\pi^{0}$}\fi}
\def\dztokpws{\relax\ifmmode{D^{0}\rightarrow K^{+}\pi^{-}}
                 \else{$D^{0}\rightarrow K^{+}\pi^{-}$}\fi}
\def\dztokppzrs{\relax\ifmmode{\overline{D^{0}}\rightarrow K^{+}\pi^{-}\pi^{0}}
                 \else{$\overline{D^{0}}\rightarrow K^{+}\pi^{-}\pi^{0}$}\fi}
\def\mkppz{\relax\ifmmode{m(K\pi\pi^{0})}
                 \else{$m(K\pi\pi^{0})$}\fi}
\def\mpkpz{\relax\ifmmode{m(\pi K\pi^{0})}
                 \else{$m(\pi K\pi^{0})$}\fi}
\def\mkkpz{\relax\ifmmode{m(KK\pi^{0})}
                 \else{$m(KK\pi^{0})$}\fi}
\def\mpppz{\relax\ifmmode{m(\pi\pi\pi^{0})}
                 \else{$m(\pi\pi\pi^{0})$}\fi}

\def\dzero{\relax\ifmmode{D^{0}}
                 \else{$D^{0}$}\fi}
\def\rws{\relax\ifmmode{R}
                 \else{$R$}\fi}
\def\mev{\relax\ifmmode{{\rm MeV}}
                 \else{MeV}\fi}
\def\mevc{\relax\ifmmode{{\rm MeV}/c}
                 \else{MeV/$c$}\fi}

\begin{document}

\preprint{\tighten\vbox{\hbox{\hfil CLNS 01/1731}
                        \hbox{\hfil CLEO 01-7}
}}

\title{Rate Measurement of \dztokppzws\ and Constraints \\ on
  \dzdzb\ Mixing}  

\author{CLEO Collaboration}
\date{\today}

\maketitle
\tighten

\begin{abstract} 
We present an observation and rate measurement of the decay 
\dztokppzws\ produced in 9~fb$^{-1}$ of $e^{+}e^{-}$
collisions near the $\Upsilon(4S)$ resonance.  
The signal is inconsistent with an upward fluctuation of the
background by \signifsig\ standard deviations.  We measured the rate of
\dztokppzws\ normalized to the rate of \dztokppzrs\ to be \rwsmeas.
This decay can be produced by doubly-Cabibbo-suppressed decays or
by the \dzero\ evolving into 
a $\overline{D^{0}}$ through mixing, followed by a Cabibbo-favored
decay to $K^{+}\pi^{-}\pi^{0}$.   
We also found the $CP$ asymmetry \asymmnumber\ to be
consistent with zero.
\end{abstract}
\newpage

{
\renewcommand{\thefootnote}{\fnsymbol{footnote}}

\begin{center}
G.~Brandenburg,$^{1}$ A.~Ershov,$^{1}$ D.~Y.-J.~Kim,$^{1}$
R.~Wilson,$^{1}$
T.~Bergfeld,$^{2}$ B.~I.~Eisenstein,$^{2}$ J.~Ernst,$^{2}$
G.~E.~Gladding,$^{2}$ G.~D.~Gollin,$^{2}$ R.~M.~Hans,$^{2}$
% ...
\end{center}
 
\small
\begin{center}
$^{1}${Harvard University, Cambridge, Massachusetts 02138}\\
$^{2}${University of Illinois, Urbana-Champaign, Illinois 61801}\\
% ...
\end{center}

\setcounter{footnote}{0}
}
\newpage


The transition of a $D^{0}$ into a $\overline{D^{0}}$ through 
mixing provides a window through which we may observe 
the effects of non-Standard Model physics.  Just as \kzkzb\ and
\bzbzb\ mixing gave prescient information about the charm and top quarks before
their discovery, observation of \dzdzb\ mixing could imply 
evidence for new
particles as massive as 100-1000 TeV~\cite{bib:Leurer}.  

In this Letter we report the first observation of the
``wrong sign'' decay $D^{0} \rightarrow K^{+}\pi^{-}\pi^{0}$
(consideration of charge-conjugate modes is 
implied throughout this Letter).  The flavor of the initial
$D^{0}$ was tagged by the sign of the slow pion, $\pi_{s}$, 
from $D^{*+} \rightarrow D^{0}\pi_{s}^{+}$.
We measured the ratio $R$ of wrong sign (WS) to right sign (RS) $D^{0}
\rightarrow K\pi\pi^{0}$ decay rates, integrated
over decay times.  The WS decays can be produced by mixing of the
initial $D^{0}$ to a $\overline{D^{0}}$, followed by Cabibbo-favored
decay (CFD) to $K^{+}\pi^{-}\pi^{0}$, or by doubly-Cabibbo-suppressed
decay (DCSD).

The transition of a $D^{0}$ to a $\overline{D^{0}}$ can proceed through
real or virtual intermediate states, which we describe by the
normalized amplitudes $-iy$ and $x$, respectively~\cite{bib:Lee}.  
The Standard Model contribution to $|x|$ is suppressed by at least
$\tan^{2}\theta_{C} \approx 0.05$ due to weak couplings, however GIM 
cancellation~\cite{bib:GIM} could further suppress $|x|$ by one to
four orders of magnitude.
While the Standard Model contributions are most likely below the present
experimental sensitivities, many non-Standard Model processes could
lead to $|x|>0.01$~\cite{bib:Harry}.    

The interference between the mixing and DCSD amplitudes is influenced
by a strong interaction phase difference between the CFD and DCSD amplitudes.
We denote the ratio of DCSD and CFD amplitudes by
$-\sqrt{R_{D}}e^{-i\overline{\delta}}$.  The leading
minus sign is motivated by the relevant
Cabibbo-Kobayashi-Maskawa matrix elements: 
${\cal R}\{ V_{cd}V^{*}_{us}/V_{cs}V^{*}_{ud} \} <0$.
An average over relevant three-body configurations of
$K^{+}\pi^{-}\pi^{0}$ is implied in $R_{D}$ and in the
strong phase $\overline{\delta}$~\cite{bib:DeltaS}.

The effect of the strong phase is to yield measurable mixing
amplitudes $y^{\prime} \equiv y\cos\overline{\delta}
-x\sin\overline{\delta}$ and 
$x^{\prime} \equiv x\cos\overline{\delta}+
y\sin\overline{\delta}$.  Then,
\[
  R \equiv \frac{\Gamma(\dztokppzws)}
  {\Gamma(\dztokppzrs)} = R_{D} + \sqrt{R_{D}}y^{\prime} 
  + \frac{1}{2}(x^{\prime 2} + y^{\prime 2})  \
   .
\]

Data for this measurement were produced in $e^{+}e^{-}$ collisions
within the CESR ring at center of mass energies near the
$\Upsilon(4S)$ resonance. Data corresponding to 9~fb$^{-1}$ of
integrated luminosity were collected using the CLEO~II.V upgrade of
the CLEO~II detector~\cite{bib:CLEOdet} between February 1996 and
February 1999.
Reconstruction of displaced decay vertices from charmed particles
was made possible by the improved resolution of the silicon vertex detector
(SVX)~\cite{bib:CLEOSVX}, installed as a part of this upgrade.
We utilized this improved resolution in previous
searches for \dzdzb\ mixing~\cite{bib:CLEOAsner} and in measurements of
charmed particle lifetimes~\cite{bib:CLEOPrell,bib:CLEOHart}.

Candidates for \dztokppzws\ were formed by combining good quality charged
tracks detected in the drift chamber with $\pi^{0}$'s
formed from pairs of photons detected in the 
CsI crystal calorimeter.  Photons from the central (end)
region of the calorimeter with energies greater than 
30 \mev\ (60 \mev) were considered.
The $\pi^{0}$ candidates were required to have
momentum greater than 340 \mevc , diphoton invariant mass within two
standard deviations of the known $\pi^{0}$ mass, good mass fit
chi-squared, and at 
least one photon detected within the central region of the calorimeter.
The tracks from the charged decay products of the $D^{0}$ candidate were
required to form a vertex with confidence level of at least 0.0001. 
The track from the $\pi_{s}$ candidate was refit with the constraint that it
pass through the intersection of the $D^{0}$ candidate direction and
the CESR beam spot. 
This refit was required to have a
confidence level of at least 0.0001.  $D^{*}$ candidates with
momentum less than 2.5 GeV/$c$ were rejected.  
We separated signal from background using distributions of
$D^{0}$ candidate mass, $M \equiv \mkppz$, and energy
released in the $D^{* +}$ decay, $Q \equiv m(K\pi\pi^{0}\pi_{s}) - \mkppz
- m_{\pi}$. 
Charged kaon and pion daughters of the \dzero\ were required to have
specific ionization consistent with their respective hypotheses.
Combinations from RS decays in which both the charged kaon and pion
were misidentified were removed by requiring the mass of the
interchanged charged track hypothesis, \mpkpz, to reconstruct at least four
standard deviations away from the known \dzero\ mass.  Similar
kinematic vetoes were applied to \mkkpz\ and \mpppz\ in order to remove
combinations with a single particle misidentification.

In this analysis, systematic uncertainties were reduced by directly fitting
for the scale factor $S$ that relates the large number of RS
events, $N_{RS}$, to the modest number of WS events, $N_{WS}$:
$N_{WS}=S\cdot N_{RS}$. 
Then, $R = C \cdot S$, where the correction $C$ can
deviate from unity because the WS events can 
populate the Dalitz plot differently than the RS events do, and thus
have a slightly different average efficiency.

The scale factor $S$ was determined using
a binned maximum likelihood fit to the two-dimensional distribution of
$Q$ and $M$.
The prominent and nearly
background-free RS signal in the data was scaled by the
factor $S$ to provide the WS signal distribution for the fit.  
Background $K\pi\pi^{0}\pi_{s}$ combinations were broken down into
three categories according to their expected distributions in $Q$ and
$M$: 1) RS \dztokppzrs\ daughters combined with an uncorrelated slow
pion, 2) charmed particle decays other than correctly reconstructed RS
\dztokppzrs , and 3) products of $e^{+}e^{-} \rightarrow u\overline{u}$,
$d\overline{d}$, or 
$s\overline{s}$ events.  Events from $b\overline{b}$ production were
excluded by the $D^{*}$ candidate momentum cut.  
The contribution from RS \dztokppzrs\
combined with an uncorrelated slow pion produces a peak in $M$, but
is smooth in $Q$.  While some backgrounds produce peaks in one
variable, none of them produce peaks in both $Q$ and $M$.
$Q$-$M$ distributions for the three backgrounds were taken from Monte
Carlo simulations, however their normalizations were allowed to vary
in the fit.  We generated approximately 40 million
$e^{+}e^{-} \rightarrow u\overline{u}$, 
$d\overline{d}$ $s\overline{s}$ and $c\overline{c}$ Monte Carlo
events, corresponding to approximately eight times the integrated luminosity
of the data, using the GEANT-based CLEO
detector simulation~\cite{bib:GEANT}. 

\begin{figure}[htp]
\centerline{
\epsfxsize=165mm
\epsffile{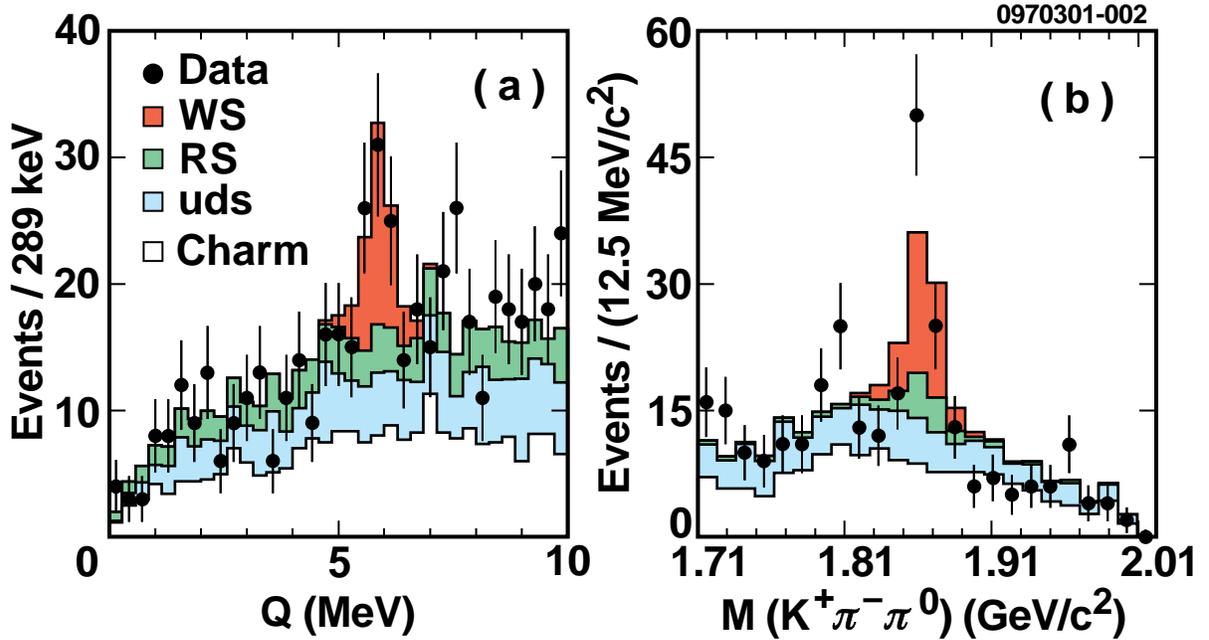}
}
\caption{Results of the $Q$-$M$ fit to the WS data, shown in 
  projections onto a) $Q$ and b) $M$
  within the signal region of two standard deviations
  about the known $M$ and $Q$ values, respectively.  Contributions to
  the WS data 
  sample come from the WS signal (WS), RS $\overline{D^{0}}$ decays combined with
  an uncorrelated slow pion (RS), decay products of $e^{+}e^{-}
  \rightarrow u\overline{u},\  d\overline{d},\ {\rm or}\
  s\overline{s}$ (uds), and decays from charmed particles, other than
  correctly reconstructed RS events (Charm).}
\label{fig:DA}
\end{figure}

The fit to the WS signal in the $Q$-$M$ plane determined a scale
factor $S$ of \rwsmeasstat\ between 
the RS and WS signal peaks.   This corresponds to a WS yield of
\nwsevnt\ events within the signal region of 
two standard deviations about the known $Q$ and $M$ values.
The statistical significance of this signal was evaluated by fitting
with the signal contribution constrained to zero and comparing the maximum
log-likelihood value with that of the nominal fit.  The signal was found to
be inconsistent with an upward fluctuation of the background by 
\signifsig\ standard deviations.  Projections of the WS signal
and fit results in the two fit variables are shown in
Fig.~\ref{fig:DA}.

The correction factor $C$ was estimated
by fitting the WS Dalitz plot. 
This correction can differ from unity if the RS and WS Dalitz
plots have different resonance structure, since the efficiency is not
uniform across the Dalitz plot.  
Recently, CLEO performed a Dalitz analysis of the RS decay
$\overline{D^{0}} \rightarrow K^{+}\pi^{-}\pi^{0}$~\cite{bib:CLEOBergfeld} 
and found a rich structure consisting  
of $\rho(770)^{-}$, $K^{*}(892)^{+}$, $K^{*}(892)^{0}$, 
$\rho(1700)^{-}$, $K_{0}(1430)^{0}$, $K_{0}(1430)^{+}$,
and $K^{*}(1680)^{+}$ resonances and non-resonant contributions.
The dominant intermediate state in the RS decay is $\overline{D^{0}}\rightarrow
K^{+}\rho(770)^{-}$, followed by $\overline{D^{0}}\rightarrow 
K^{*}(892)^{+}\pi^{-}$ and  
$\overline{D^{0}}\rightarrow K^{*}(892)^{0}\pi^{0}$, which account
for roughly \rhopctyld, \ksppctyld, and \kszpctyld\ of the yield, 
respectively. The sum of these is greater than 100\% due to 
interference.

We used an unbinned maximum likelihood fit to extract the relative
contributions of the three major resonances from the WS data.  Due to
the limited statistics and relatively large background,  
only the amplitudes and phases of $D^{0}\rightarrow 
K^{*}(892)^{+}\pi^{-}$ and
$D^{0}\rightarrow K^{*}(892)^{0}\pi^{0}$ relative to
$D^{0}\rightarrow K^{+}\rho(770)^{-}$ were varied in the fit.  Relative
amplitudes and phases of the minor contributions were fixed to the RS
values found in~\cite{bib:CLEOBergfeld}.   

The fit used the signal fraction from the $Q$-$M$ fit and
parameterizations of the efficiency and background distributions in
the Dalitz plot variables 
as inputs.  An analytic expression for the efficiency function 
was obtained by fitting a large sample of non-resonant signal
\dztokppzws\ Monte Carlo events.
The background function was estimated by fitting side-band regions in $Q$ 
from the WS data.  These regions contain contributions from
RS \dztokppzrs\ combined with an uncorrelated slow pion, $e^{+}e^{-}
\rightarrow u\overline{u}$, $d\overline{d}$, 
$s\overline{s}$, and non-signal charm events.  
The measured RS squared amplitude~\cite{bib:CLEOBergfeld}, 
multiplied by the efficiency function, was used to describe the Dalitz
plot of the RS \dztokppzrs\ with uncorrelated $\pi_{s}$ 
contribution.  The combined $e^{+}e^{-} \rightarrow u\overline{u}$,
$d\overline{d}$, or $s\overline{s}$ and non-signal charm fit function
was taken to be a two-dimensional polynomial with coefficients
determined from the $Q$ side band fit.  The relative contributions of
these backgrounds were obtained from the $Q$-$M$ fit result.

The results of this fit were surprisingly consistent with those of the RS
fit~\cite{bib:CLEOBergfeld},
leading to a correction of $C=1.00 \pm 0.02\ ({\rm stat})$.  While the
statistical uncertainties on the amplitudes and phases were
large due to the low statistics and large backgrounds,
the correction factor $C$ was insensitive to these because
of the relatively uniform efficiency.  This resulted in a
statistical uncertainty on $C$ that was small compared with that for $S$.

The total systematic uncertainty on $R$ was estimated to be
\totalsyserr .  Contributions to this uncertainty are categorized in
Tab.~\ref{tab:errors} as measurement errors on $S$ and $C$ from the
$Q$-$M$ and Dalitz plot fits, respectively.  

The important systematic uncertainties on $S$ were due to uncertainties in
the Monte Carlo simulation of the background $Q$-$M$ distributions
(\errQmshapes), sensitivity to misidentification of charged $D^{0}$
daughter tracks (\mccutdisterr), and the 
statistics of the Monte Carlo sample (\errmcstat).  These are
summarized in Tab.~\ref{tab:errors}.
The largest of these, due to the simulation of the background $Q$-$M$
distributions used in the fit, was estimated
by varying the $Q$-$M$ side band regions used to constrain
the background contributions.  
The sensitivity to misidentification of the daughter $K^{+}$ and
$\pi^{-}$ was studied by observing the variation of $S$ with changes
in the cuts applied to specific ionization and the kinematic cuts
applied to \mpkpz , \mkkpz , and \mpppz .  
The uncertainty due to the statistics of the Monte
Carlo background $Q$-$M$ distributions was estimated by performing a
series of fits in which the contents of the bins were varied according
to Poisson statistics.  The uncertainty was taken to be the standard
deviation of \rws\ based on 4000 variations.

Uncertainties in the measurement of $C$ come from the unknown
minor resonance contributions to the Dalitz plot (\errotherres), 
the Dalitz plot fit method (\onehyperr), the unknown background Dalitz
plot shapes (\errdalbgdst), and the statistical uncertainty in the Dalitz
plot fit (\errdalfitstat).  These contributions are summarized
in Tab.~\ref{tab:errors}. 
The amplitudes and phases of resonant and non-resonant components 
of the WS signal other than the dominant $K^{+}\rho(770)^{-}$,
$K^{*}(892)^{+}\pi^{-}$, and $K^{*}(892)^{0}\pi^{0}$ modes were
fixed to the RS values in the fit.  In order to explore the uncertainty
of this assumption, we allowed these to vary in the fit.
The systematic uncertainty in the Dalitz plot fit method was
determined by fitting the WS Dalitz plot under hypotheses that the
signal was composed entirely of $K^{+}\rho(770)^{-}$, 
$K^{*}(892)^{+}\pi^{-}$, or $K^{*}(892)^{0}\pi^{0}$ decays.
$C$ was found to differ 
from one by \allrhoerr, \allksperr, and \allkszerr, respectively.  The
uncertainty was estimated by evaluating the consistency between the
WS data and the pure $K^{*}(892)^{0}\pi^{0}$ 
hypothesis, which produced the largest deviation from unity.  This hypothesis
was found to be inconsistent with the data by 7.7 standard deviations.  
The systematic uncertainty due to the Dalitz plot of the background was
estimated using a series of fits with background shapes from side
bands in $Q$ and  
$M$, obtained from both the WS and RS data.  When using
side bands in $M$, the kinematics of the daughter tracks of
$D^{0}$ candidates were scaled to
force the allowed phase space to be similar to that of a true
$D^{0}\rightarrow K\pi\pi^{0}$ decay.
The statistical uncertainty on $C$ from the Dalitz
plot fit was included as a systematic uncertainty on \rws. 
The Dalitz plot fit results were checked by performing $Q$-$M$ and $Q$ fits to
specific Dalitz plot subregions dominated by a single submode.  The
relative yields from these fits were compared and found to be in
agreement with the efficiency-corrected squared amplitude from the
Dalitz plot fit, integrated over the same subregion.  The total systematic
uncertainty on $C$ was estimated to be \effratiototerr .

By performing the analysis separately for $D^{0}$ and
$\overline{D^{0}}$ candidates, we measured the $CP$ asymmetry of this
decay, defined to be 
\[
  A=\frac{R(D^{0}\rightarrow K^{+}\pi^{-}\pi^{0}) - 
  R(\overline{D}^{0}\rightarrow K^{-}\pi^{+}\pi^{0})} 
  {R(D^{0}\rightarrow K^{+}\pi^{-}\pi^{0}) + 
  R(\overline{D}^{0}\rightarrow K^{-}\pi^{+}\pi^{0})} .
\]
We observed an asymmetry consistent with zero: \asymmnumber .
Due to cancellation of errors in this ratio, the
systematic uncertainty in this measurement was negligible
compared with its statistical uncertainty.

In summary, we observed a signal for the decay \dztokppzws\ using
9~fb$^{-1}$ data collected with the CLEO~II.V detector.  The signal is 
inconsistent with an upward fluctuation of the background by
\signifsig\ standard deviations.  This result is the first observation
of the WS signal $D^{0}\rightarrow K^{+}\pi^{-}\pi^{0}$. 
Using fits to the $Q$-$M$ and Dalitz plots, we measured
the normalized WS rate and $CP$ asymmetry to be R = \rwsmeas\ and
\asymmnumber , respectively. 

\begin{figure}[htp]
\centerline{
\epsfxsize=165mm
\epsffile{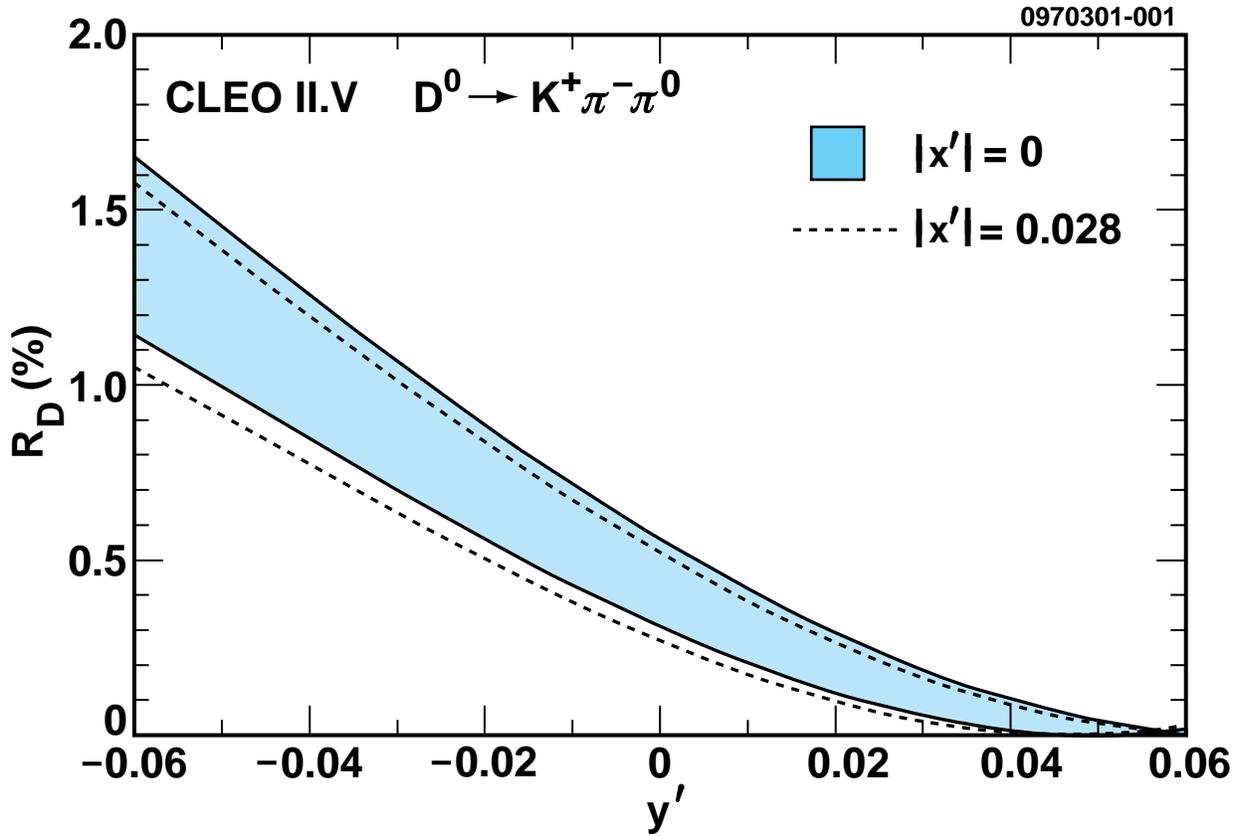}
}
\caption{Doubly-Cabibbo-suppressed rate as a function of 
$y^{\prime}$, plotted for two values of $|x^{\prime}|$ which
correspond to the upper and lower limits from the CLEO
$D^{0}\rightarrow K^{+}\pi^{-}$ measurement 
if equal strong interaction phase differences are assumed for the two modes.
The bands indicate the region within one standard deviation of this
measurement.}  
\label{fig:rwsmeas}
\end{figure}

To allow comparison with previous measurements of DCSD and mixing in the $D^0$
system~\cite{bib:CLEOAsner,bib:CLEOPrescott,bib:FOCUSKpi,bib:FOCUSKK} we
plot a band corresponding to this measurement of 
$R$ in the $R_{D}$--$y^{\prime}$ plane, shown in
Fig.~\ref{fig:rwsmeas}. 
The band depends on $|x^{\prime}|$ and we show it for
$|x^{\prime}| = 0$ and $|x^{\prime}| = 
0.028$, which correspond to the limits from our previous analysis of 
$D^0$ decay to $K^{+}\pi^{-}$~\cite{bib:CLEOAsner} if equal strong interaction
phase differences are assumed for the decays~\cite{bib:DeltaS}.  If we assume
that there is no mixing, this measurement corresponds to
$R_{D}=(1.7\pm 0.4\ ({\rm   stat}) \pm 0.3\ ({\rm syst}))\cdot
\tan^{4}\theta_{C}$. 


We gratefully acknowledge the effort of the CESR staff in providing us with
excellent luminosity and running conditions.
This work was supported by 
the National Science Foundation,
the U.S. Department of Energy,
the Research Corporation,
the Natural Sciences and Engineering Research Council of Canada
and the Texas Advanced Research Program.

\begin{table}
\caption[]{Systematic uncertainties in the $R$ measurement.}
\label{tab:errors}
\begin{center}
\begin{tabular}{|l|l|r|}
Measurement & Source  &  Uncertainty  \\ \hline  
$S$ & MC background $Q$-$M$ dist. & \errQmshapes \\ 
    & $K/\pi$ separation & \mccutdisterr  \\  
    & MC statistics of $Q$-$M$ dist. & \errmcstat \\ 
\hline
$C$ & Minor resonances & $8\%$ \\
    & Dalitz fit method & $3.6\%$ \\
    & Background Dalitz plot & $3\%$ \\
    & Dalitz fit stat. error & $2\%$ \\
\hline
$R=C\cdot S$    &  & \totalsyserr  \\
\end{tabular}
\end{center}
\end{table}

\begin{thebibliography}{99}


\bibitem{bib:Leurer} M. Leurer, Y. Nir, and N. Seiberg, 
  Nucl. Phys. B {\bf 420}, 468 (1994); N. Arkani-Hamed, {\em et al.},
  Phys. Rev. D {\bf 61}, 116003 (2000).  
\bibitem{bib:Lee} T. D. Lee, R. Oehme, and C. N. Yang, 
  Phys. Rev. {\bf 106}, 340 (1957); A. Pais and S. B. Trieman, 
  Phys. Rev. D {\bf 12}, 2744 (1975).
\bibitem{bib:GIM} S. L. Glashow, J. Iliopolous, and L. Maiani,
  Phys. Rev. D {\bf 2}, 1285 (1970).  
\bibitem{bib:Harry} H. N. Nelson, hep-ex/9908021 (unpublished).
\bibitem{bib:DeltaS} The variables $\overline{\delta}$, $x^{\prime}$,
  $y^{\prime}$, and $R_{D}$ presented in this paper represent averages over
  all submodes contributing to the $K^{+}\pi^{-}\pi^{0}$ final state.  
  Comparison with the corresponding measured variables in the
  $D^{0} \rightarrow K^{+}\pi^{-}$ mode cannot be made without
  assumptions about the relative phases of final state interactions.
\bibitem{bib:CLEOdet}  Y. Kubota {\em et al.}, CLEO Collaboration,
  Nucl. Instrum. Meth. A {\bf 320}, 66 (1992).
\bibitem{bib:CLEOSVX} T. S. Hill,
  Nucl. Instrum. Meth. A {\bf 418}, 32 (1998).
\bibitem{bib:CLEOAsner} R. Godang {\em et al.}, CLEO Collaboration,
  Phys. Rev. Lett. {\bf 84}, 5038 (2001).
\bibitem{bib:CLEOPrell} G. Bonvicini {\em et al.}, CLEO Collaboration,
  Phys. Rev. Lett. {\bf 82}, 4586 (1999).
\bibitem{bib:CLEOHart} A. H. Mahmood {\em et al.}, CLEO Collaboration,
  Phys. Rev. Lett. {\bf 86}, 2232 (2001).
\bibitem{bib:GEANT} ``QQ--The CLEO Event Generator,''
  http:$//$www.lns.cornell.edu$/$public$/$CLEO$/$soft$/$QQ
  (unpublished); T. Sj\o strand, Comput. Phys. Commun., {\bf 39}, 347
  (1986); T. Sj\o strand and M. Bengston, Comput. Phys. Commun. {\bf
  43}, 367 (1987);  R. Brun {\it et al.}, CERN Report No. DD/EE/84-1 (1987).
\bibitem{bib:CLEOBergfeld} S. Kopp {\em et al.}, CLEO Collaboration,
  Phys. Rev. D {\bf 63}, 092001 (2001).
\bibitem{bib:CLEOPrescott} A. B. Smith, in {\em 
       Proceedings of the Fourth International Conference on B Physics
       and CP Violation}, Ise-Shima, February 2001 (to be published); 
      hep-ex/0104008.
\bibitem{bib:FOCUSKpi} J. M. Link {\em et al.}, FOCUS Collaboration, 
Phys. Rev. Lett. {\bf 86}, 2955 (2001).
\bibitem{bib:FOCUSKK} J. M. Link {\em et al.}, FOCUS Collaboration,
  Phys. Lett. B {\bf 485}, 62 (2000).

\end{thebibliography}
\end{document}